\documentclass[aps,prl,preprintnumbers,amsmath,amssymb,
               superscriptaddress,groupedaddress,nofootinbib,
               tightenlines,twocolumn]{revtex4-1}

\usepackage{graphicx}
\usepackage{sidecap}
\usepackage{amsmath}
\usepackage{bbold}
\usepackage{bm}
\usepackage{bbm}
\usepackage{color}
\usepackage{soul}
\usepackage{diagbox}
\usepackage{etoolbox}
\apptocmd{\sloppy}{\hbadness 10000\relax}{}{}
\usepackage[bookmarks=false]{hyperref}

\def\beqn{\begin{eqnarray}}
\def\eeqn{\end{eqnarray}}
\def\barr{\begin{array}}
\def\earr{\end{array}}
\def\btab{\begin{tabular}}
\def\etab{\end{tabular}}
\def\bite{\begin{itemize}}
\def\eite{\end{itemize}}
\def\bcen{\begin{center}}
\def\ecen{\end{center}}

\def\eq{\begin{equation}}
\def\ee{\end{equation}}
\def\nn{\nonumber}

\def\q2dagger{q_2\hspace{-0.35cm}/\;}

\newcommand{\rhoX}[1]{{\rho}_{\raisebox{-2.50pt}{\!\tiny #1}}}
\newcommand{\be}{\begin{equation}}
\newcommand{\bea}{\begin{eqnarray}}
\newcommand{\eea}{\end{eqnarray}}

\begin{document}
\preprint{MITP/20-020}
\title{
Weak charge and weak radius of ${}^{12}$C
}

\author{Oleksandr Koshchii$^{a}$}
\email[]{koshchii@uni-mainz.de}
\author{Jens Erler$^{a,b,c}$}
\author{Mikhail Gorchtein$^{d}$}
\author{Charles J. Horowitz$^{e}$}
\author{Jorge~Piekarewicz$^{f}$}
\author{Xavier Roca-Maza$^{g}$}
\author{Chien-Yeah Seng$^{h}$}
\author{Hubert Spiesberger$^{a}$}

\affiliation{$^{a}$PRISMA$^+$Cluster of Excellence,
    Institut f\"ur Physik,\\
	Johannes Gutenberg-Universit\"at, D-55099 Mainz, Germany}
\affiliation{$^{b}$Helmholtz Institute Mainz,
	Johannes Gutenberg-Universit\"at, D-55099 Mainz, Germany}
\affiliation{$^{c}$Departamento de F\'{\i}sica Te\'{o}rica,
    Instituto de F\'{\i}sica,\\
	Universidad Nacional Aut\'{o}noma de M\'{e}xico, 04510 CDMX,
	M\'{e}xico}
\affiliation{$^{d}$PRISMA$^+$Cluster of Excellence,
    Institut f\"ur Kernphysik,\\
	Johannes Gutenberg-Universit\"at, D-55099 Mainz, Germany}
\affiliation{$^{e}$Center for Exploration of Energy and Matter
    and Department of Physics, Indiana University, Bloomington,
    Indiana 47405, USA}
\affiliation{$^{f}$Department of Physics, Florida State University,
    Tallahassee, Florida 32306, USA}
\affiliation{$^{g}$Dipartimento di Fisica, Università degli Studi
    di Milano, Via Celoria 16, I-20133 Milano, Italy \\
    and INFN, Sezione di Milano, Via Celoria 16, I-20133 Milano,
    Italy}
\affiliation{$^{h}$Helmholtz-Institut f\"ur Strahlen- und Kernphysik
    and Bethe Center for Theoretical Physics,\\ Universit\"at Bonn,
    D-53115 Bonn, Germany}


\begin{abstract}
We present a feasibility study of a simultaneous sub-percent
extraction of the weak charge and the weak radius of the
${}^{12}$C nucleus using parity-violating electron scattering, based
on a {largely} model-independent assessment of the uncertainties.
The corresponding measurement is considered to be carried out
at the future MESA facility in Mainz with $E_{\rm beam} = 155$ MeV.
We find that a combination of a 0.3\%
precise measurement of the parity-violating asymmetry at forward
angles with a 10\% measurement at backward angles will allow to
determine the weak charge and the weak radius of ${}^{12}$C with
0.4\% and 0.5\% precision, respectively. These values could be
improved to 0.3\% and 0.2\% for a 3\% backward measurement.
This experimental program will have impact on precision low-energy
tests in the electroweak sector and nuclear structure.
\end{abstract}
\date{\today}
\maketitle


Precise measurements of the parameters of the standard model
(SM) are among the main tools to search for or constrain
hypothetical contributions from physics beyond the SM. The
central parameter of the electroweak sector of the SM is the
weak mixing angle $\theta_W$ describing the mixing of the $\rm{SU}(2)$
and $\rm{U}(1)$ gauge boson fields, which results in the emergence of
the physical fields, the massless photon and the massive $Z^0$.
Its sine squared, $\sin^2\theta_W$, is related to the vector
charge of SM fermions with respect to the weak neutral current
and can be accessed in various processes and at different
energy scales: from $Z$-pole measurements at colliders
\cite{ALEPH:2005ab,Aaltonen:2018dxj}, including the LHCb
\cite{Aaij:2015lka}, ATLAS \cite{atlas} and CMS
\cite{Sirunyan:2018swq} experiments, to deep inelastic
scattering with electrons \cite{Prescott:1997,Wang:2014} and
neutrinos \cite{Zeller:2001hh}, to parity violation in atoms
\cite{Wood:1997,Guena:2004sq} and to parity-violating electron
scattering (PVES) off protons \cite{qweak} and electrons
\cite{Anthony:2005pm}. To connect these measurements across
the relevant energy scales, the SM running at the one-loop level
needs to be taken into account \cite{Erler:2004in,Erler:2017knj}.
Currently, this running is theoretically known at the relative
level of $\sim8\times10^{-5}$, which provides the basis for an
ambitious experimental program at low energies: an ongoing effort
in atomic parity violation \cite{Zhang:2016czr,Antypas:2018mxf}
has the goal to measure the weak charges of heavy nuclei and
chains of nuclear isotopes at the per mille precision. The
Qweak experiment \cite{qweak} has recently extracted
$\sin^2\theta_W$ from low-energy PVES to 0.5\% accuracy.
P2@MESA \cite{Becker:2018ggl} and the MOLLER Collaboration
\cite{Benesch:2014bas} aim at improving that result by factors
of 4 and 6, respectively. Further plans involve deep-inelastic
electron scattering with SOLID \cite{solid}.

Apart from tests of the SM, PVES has also been used to address
aspects of  nucleon and nuclear structure that are elusive to
photons. PVES off heavy nuclei with a neutron excess is used to
determine the neutron skin \cite{Thiel:2019tkm}---the difference
in the radii of the neutron and proton distributions---with the
goal of constraining the equation of state (EOS) of neutron rich
matter\,\cite{Horowitz:2000xj}. The lead (Pb) Radius EXperiment
(PREX) \cite{Abrahamyan:2012gp} has provided the first
model-independent evidence in favor of a neutron-rich skin in
${}^{208}$Pb \cite{Horowitz:2012tj}. Further experiments with
an improved precision are presently being analyzed \cite{PREX-II},
running \cite{CREX}, or planned \cite{Becker:2018ggl}. PVES off
the proton and light nuclei has been extensively used to determine
the strange quark content of the nucleon \cite{Maas:2017snj}.

\begin{figure}[t]
  \centering
  \resizebox{0.3\textwidth}{!}{\includegraphics{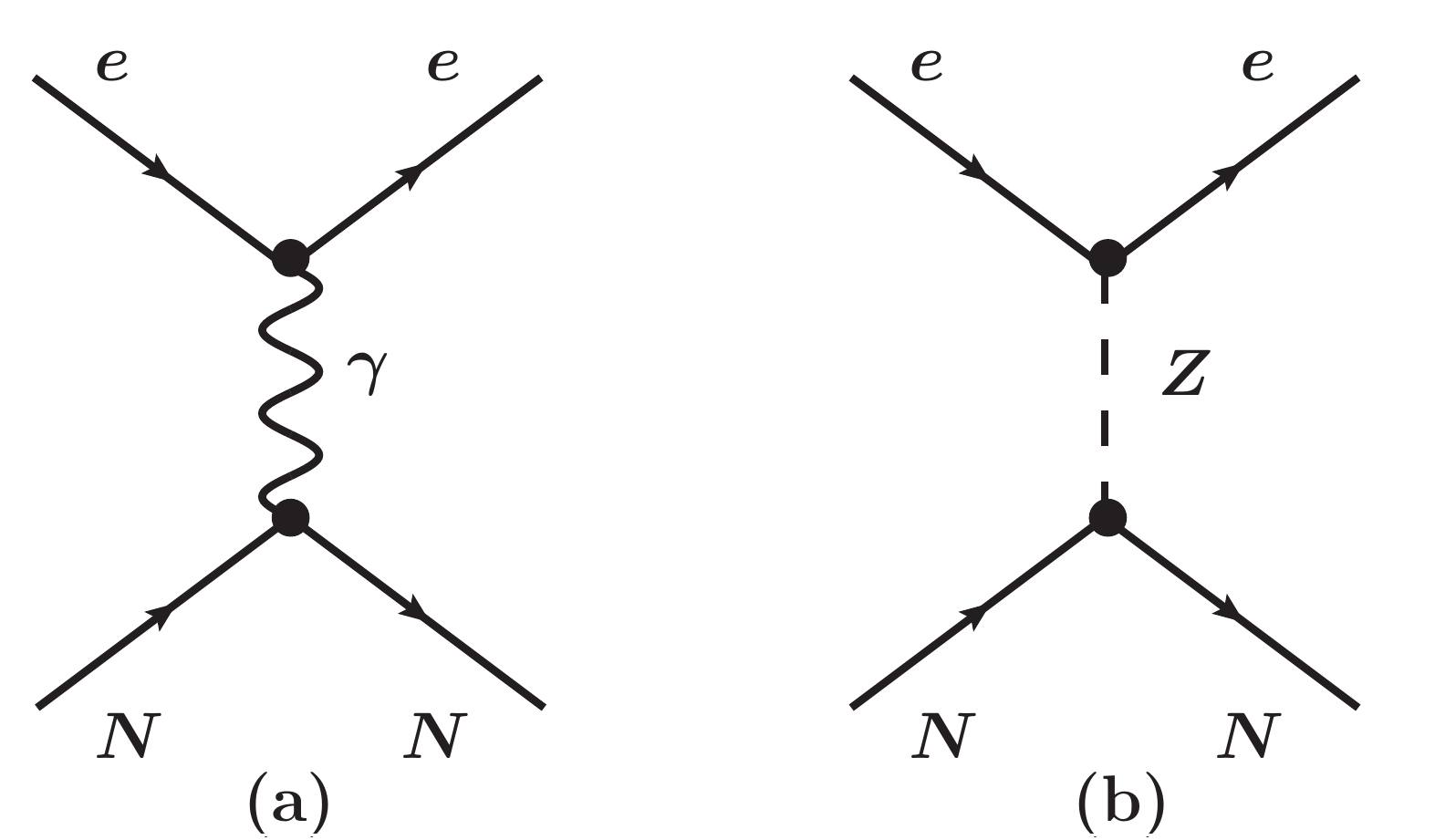}}
  \caption{(a) One-photon exchange and (b) $Z^{0}$ exchange diagrams.
  }
  \label{fig:tree}
\end{figure}

In this Rapid Communication, we consider the parity-violating asymmetry
which is defined as the difference between the cross sections for elastic
scattering of {longitudinally} polarized electrons off an
unpolarized target,
\beqn
A^{\rm PV}=\frac{\sigma_R-\sigma_L}{\sigma_R+\sigma_L},
\eeqn
where $\sigma_{R} \, (\sigma_L)$ stands for the cross section with
right-handed (left-handed) electron polarization. The asymmetry
arises from the interference between the amplitudes due to the
exchange of a virtual photon and {the corresponding one} for a
virtual $Z^{0}$ boson as shown in Fig.~\ref{fig:tree}.
By conveniently factoring out the Fermi constant $G_F$, the 
fine-structure constant $\alpha$, the four-momentum transfer squared
$Q^2$, and the ratio of the weak, $Q_W$, to the electric, $Z$,
nuclear charge, the PV asymmetry {for a spinless}
nucleus consisting of $Z$ protons and $N$ neutrons takes the
following form:
\beqn
A^{\rm PV}
=
- \frac{G_FQ^2}{4\sqrt2\pi\alpha} \frac{Q_W}{Z}(1+\Delta),
\label{eq:APV}
\eeqn
where a plane-wave Born (``tree-level") approximation was
assumed. The weak nuclear
charge is given by $Q_W(Z,N)\!=\!Z(1\!-\!4\sin^2\theta_W)\!-\!N$,
so, in the case of ${}^{12}$C, it becomes proportional to the
sine squared of the weak mixing angle \cite{ftnt1}:
$Q_W(6,6)\!=\!-24\sin^2\theta_W$.

Given that the interaction of the electron with the nucleus
involves only the conserved hadronic vector current,
the ``correction" term $\Delta$ in Eq.~(\ref{eq:APV}) vanishes
at $Q^2\!=\!0$. However, nuclear and hadronic
structures contribute to $\Delta$ at non zero $Q^2$.
Indeed, to leading order in $\alpha$,
\beqn
\Delta \equiv F_{\rm wk}(Q^2)/F_{\rm ch}(Q^2) - 1
\label{eq:DeltaDef}
\eeqn
is given by the ratio of the weak $F_{\rm wk}$ to the charge
form factor $F_{\rm ch}$. Both form factors are normalized to
unity at $Q^2\!=\!0$. Each of the form factors is related to
the corresponding spatial distributions of charge by a
three-dimensional Fourier transform,
\beqn
F(Q^2) = \int\!\rho(r) e^{i{\bf q}\cdot{\bf r}} d^{3}r,
\hspace{5pt} {\rm with} \hspace{3pt} |{\bf q}|\!\equiv\!\sqrt{Q^{2}}.
\eeqn
Note that the normalization of the form factor at
$Q^2\!=\!0$ implies that $\!\int\!\rho(r) d^{3}r\!=\!1$.
At low $Q^2$, the form factors may be expanded in
terms of various moments of their spatial distribution,
\beqn
F(Q^2) = 1 - \frac{Q^{2}}{3!}\langle r^{2}\rangle +
\frac{Q^{4}}{5!}\langle r^{4}\rangle  +
{\cal O}(Q^{6}),
\eeqn
where the second term defines the root-mean-square radius of the
spatial distribution, namely,
\beqn
 R^{2} \equiv \langle r^{2} \rangle = \int r^{2} \rho(r) d^{3}r.
\label{eq:moments}
\eeqn
Thus, to lowest order in $Q^2$, $\Delta$ is
proportional to the weak skin of the nucleus,
\beqn
\Delta = - \frac{Q^{2}}{3} R_{\rm wskin} R_{\rm ch} + {\cal O}(Q^2 R_{\rm wskin}^2).
\label{eq:delta_taylor}
\eeqn
The weak skin $R_{\rm wskin}\!\equiv\!R_{\rm wk}\!-\!R_{\rm ch}$,
or, equivalently,
\beqn
\lambda = \frac{R_{\rm wk}-R_{\rm ch}}{R_{\rm ch}}
\label{eq:lambda_def}
\eeqn
coincides with the neutron skin in the idealized nonrelativistic picture,
but is distinct from it once the small weak charge of the proton,
relativistic effects, including strangeness, and radiative corrections
are taken into account.

Two terms in Eq.~(\ref{eq:APV}) are of great interest: the
weak mixing angle $\theta_W$ encoded in the weak charge
\cite{qweak,Becker:2018ggl} and the ratio of nuclear form factors
appearing in $\Delta$; to access the former, one must constrain
the latter. Conversely, to extract nucleon- or nuclear-structure
information from PVES, such as the strange quark content of
the nucleon \cite{Maas:2017snj} or the weak skin of heavy nuclei
\cite{Horowitz:2012tj}, one assumes that $Q_W$ is precisely known,
so the measurement provides a constraint on $\Delta$. In this Rapid
Communication, we explore the possibility of a precise determination 
of both---the weak charge and the weak skin of ${}^{12}$C---within one 
single experiment.

The P2 experimental program at the MESA facility in Mainz
\cite{Becker:2018ggl} includes a plan aiming for a $0.3\%$
determination of the weak charge of ${}^{12}$C. Given this
ambitious goal, the tree-level formula of
Eq.~(\ref{eq:delta_taylor})---even when including higher-order
terms in the $Q^2$ expansion---is not accurate enough.
Order-$\alpha$ radiative corrections, particularly
Coulomb distortions which scale as $Z\alpha$, should be included.
To properly account for Coulomb distortions, we follow the
formalism developed by one of us in Ref.\,\cite{Horowitz:1998vv}.
The electron wave function $\Psi$ satisfies the Dirac equation
\beqn
\Big(\bm{\alpha}\cdot {\bf p} + \beta m + V(r) + \gamma_{5}A(r)\Big)\Psi({\bf r})
= E\Psi({\bf r}),
\label{eq:dirac}
\eeqn
where $m$ is the electron mass, $\bm{\alpha}$, $\beta$, and
$ \gamma_{5}$ are Dirac matrices, and $V(r)$ and $A(r)$ are
the vector (Coulomb) and axial-vector components of the
potential, respectively \cite{Horowitz:1998vv}. Here, $E$
stands for the electron energy in the center-of-mass frame
\cite{ftnt2} which is related (neglecting the electron mass)
to the laboratory energy $E_{\rm beam}$ by
$E_{\rm beam}/E\!=\!\sqrt{1\!+\!2E_{\rm beam}/M}$ with $M$ the
nuclear mass.

The Coulomb potential is computed from the experimentally
known nuclear charge distribution via
\beqn
V(r) =
-Z\alpha\!\!\int \frac{\rho_{\rm ch}(r')}{|{\bf r}-{\bf r}'|}d^{3}r'.
\eeqn
The axial-vector potential for a point like weak charge is short-range 
$A(r)\!\propto\!\delta^3({\bf r})$, but acquires a finite
range due to the finite size of the nuclear weak charge
distribution. That is,
\beqn
A(r) = \frac{G_{F}Q_{W}}{2\sqrt2}\rho_{\rm wk}(r).
\eeqn
The Dirac equation displayed in Eq.~(\ref{eq:dirac}) is solved
numerically using the ELSEPA package \cite{ELSEPA}, properly
modified to include the axial-vector
potential\,\cite{RocaMaza:2011pm,upcoming}. The
intrinsic relative precision of the computation of $A^{\rm PV}$
is on the order of $10^{-4}-10^{-5}$, and in a calculation at the
per mille level there are only genuine uncertainties of
$\rho_{\rm wk}$ itself.

For the nuclear charge distribution of ${}^{12}$C, we use the
parametrization of the world data on elastic electron-carbon
scattering in the form of a sum of Gaussians \cite{DeJager:1987qc}.
The fact that the charge density of ${}^{12}$C and its charge
radius $R_{\rm ch}=2.4702(22)$\,fm \cite{Angeli:2013} are known
with high precision serves as the basis for an accurate extraction
of $\sin^2\theta_W$ and of $R_{\rm wk}$ from a measured
$A^{\rm PV}$. A possible avenue is to rely on models to produce
a range of predictions for $\rho_{\rm wk}$ which is then used
to directly fit the PV asymmetry to determine the value of the
weak radius as was performed in the case of the PREX. One choice 
for parametrizing the weak charge distribution is the two-parameter
symmetrized Fermi distribution,
\begin{align}
\rho_{\rm wk}(r)
= \rhoX{SF}(r,c,a)
& = \,\rhoX{0}\,\frac{\sinh(c/a)}{\cosh{(r/a)}+\cosh(c/a)},
\nonumber \\
\rhoX{0}
& = \frac{3}{4\pi c\left(c^{2}+\pi^{2}a^{2}\right)},
\label{eq:RhoSF}
\end{align}
with $c$ and $a$ as the half-density radius and surface diffuseness,
respectively, and $\rhoX{SF}$ is normalized to unity. The
advantage of the symmetrized Fermi parametrization, apart from
its simplicity, is that its form factor, and all of its moments
are known analytically \cite{Piekarewicz:2016vbn}. In particular,
the mean-square radius of the distribution is
\beqn
 R_{\rm SF}^2 = \frac{3}{5}c^{2} + \frac{7}{5} \pi^{2}a^{2}.
\eeqn

\begin{figure}[t]
  \centering
  \resizebox{0.48\textwidth}{!}{\includegraphics{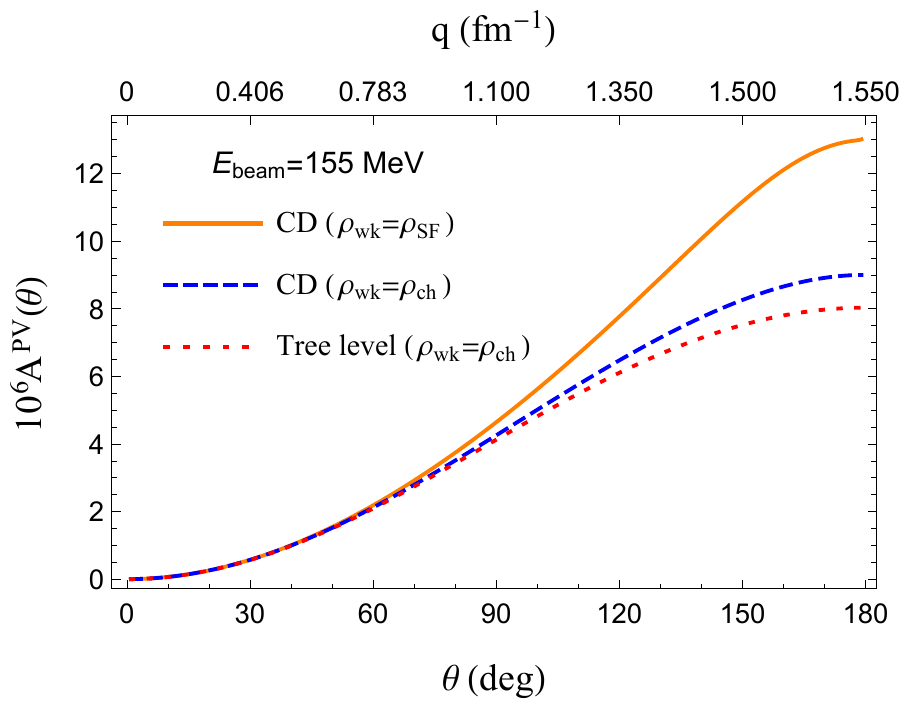}}
  \caption{
  The PV asymmetry for elastic electron scattering off ${}^{12}$C
  at $E_{\rm beam}\!=\!155$ MeV as a function of the scattering angle
  $\theta$ (lower $x$ axis) and of the momentum transfer
  $q\!=\!\sqrt{Q^2}$ (upper $x$ axis).
  We show the plane-wave (tree-level) result with
  $\rho_{\rm wk}\!=\!\rho_{\rm ch}$ (red dotted curve) and the Coulomb distorted (CD)
  predictions with $\rho_{\rm wk}\!=\!\rho_{\rm ch}$ (blue dotted curve) and with
  $\rho_{\rm wk}\!=\!\rho_{\rm SF} (r, c\!=\!2.07\,\mathrm{fm}, a\!=\!0.494\,\mathrm{fm})$
  (solid orange curve).
  }
  \label{fig:CDvsTL}
\end{figure}

In Fig.~\ref{fig:CDvsTL}, we show results for the PV asymmetry
at a fixed electron beam energy of $E_{\rm beam}\!=\!155$\,MeV
as a function of the laboratory scattering angle $\theta$ and
momentum transfer $q$. Results are displayed in both a plane-wave
(tree-level) approximation and with Coulomb distortions. The two
distorted-wave calculations
use $\rho_{\rm wk}\!=\!\rho_{\rm ch}$ (no skin) and
$\rho_{\rm wk}\!=\!\rho_{\rm SF} (r, c\!=\!2.07\,\mathrm{fm},
a\!=\!0.494\,\mathrm{fm})$ with $R_{\rm wk}\!=\!2.44\,\mathrm{fm}$
which falls within the range of values of a representative set of
nuclear-structure models \cite{Chen:2014sca,
Chen:2014mza,RocaMaza:2012sj}.
We observe a strong dependence of the PV asymmetry on the
value of the weak skin, especially at backward angles. We
also find that it is important to include effects due to
Coulomb distortions.
Our results displayed in Fig.~\ref{fig:CDvsTL} are qualitatively
similar to those obtained in Ref.~\cite{Moreno:2013pna}, but
a quantitative comparison is difficult
because of different perspectives adopted in the calculation of
the weak charge density, and several kinematic approximations
used in that paper.

Unfortunately, the choice of a particular form for the weak
charge distribution introduces model dependence that may be
difficult to quantify when extracting weak charge and radius
from a measurement of $A^{\rm PV}$: models that predict different
values for the weak radius would generally differ in all the
higher moments of the weak charge distribution, as well. To
unambiguously disentangle the effect of the weak skin, we propose
a different method. Given the $N\!=\!Z$ character of ${}^{12}$C,
its weak charge distribution is expected to follow closely the
electric charge distribution. We introduce the small difference
between the two, the ``weak-skin" distribution,
\beqn
    \rho_{\rm wskin}(r)
    \!\equiv\!
    \rho_{\rm wk}(r)\!-\!\rho_{\rm ch}(r) .
\label{rhowk}
\eeqn
Note that $\rho_{\rm wskin}$ is the spatial analog of the
weak-skin form factor depicted in Figs.~3 and 6 of
Ref.~\cite{Thiel:2019tkm}. $\rho_{\rm wskin}(r)$ is normalized
to zero and its second moment can be fixed to
\beqn
    \int \rho_{\rm wskin}(r) r^{2} d^{3}r
    =
    R_{\rm wk}^{2} \!-\! R_{\rm ch}^{2}
    =
    2 \lambda R_{\rm ch}^2 + {\cal O}(\lambda^2). \ \ \
\label{eq:skinmom}
\eeqn
This allows us to write
\beqn
    \rho_{\rm wskin}(r)
    =
    \lambda \bar{\rho}(r; \zeta) ,
\label{eq:rhobar}
\eeqn
where $\zeta$ is representative of the model dependence.
This parametrization is advantageous because it allows to
explicitly separate the dependence on $\lambda$ from the
effects of the higher moments of the weak charge density
encapsulated in a (set of) model parameter(s) $\zeta$. For
example, assuming the symmetrized Fermi parametrization of
$\rho_{\rm wk}$ as in Eq.\ (\ref{eq:RhoSF}), one would find
\beqn
&&\rho_{\rm wskin}(r)=
({\lambda}/{\lambda_{\rm SF}})\Big(\rhoX{SF}(r,c,a)
- \rho_{\rm ch}(r) \Big)
\label{eq:rhoskin}
\\
&&{\rm with} \quad
\lambda_{\rm SF}
=\lambda_{\rm SF}(c,a)
=R_{\rm SF}(c,a)/R_{\rm ch}-1 .
\nn
\eeqn
This parametrization corresponds to rewriting $\Delta$ in
Eq.~(\ref{eq:DeltaDef}) as
\begin{align}
\Delta
& = - \frac{\lambda}{3}Q^{2}R_{\rm ch}^2
+ \left(\frac{F_{\rm wk}}{F_{\rm ch}}-1
+ \frac{\lambda}{3}Q^{2}R_{\rm ch}^2\right) \nonumber \\
& = - \frac{\lambda}{3}Q^{2}R_{\rm ch}^2
+ \left[\frac{\lambda}{\lambda_{\rm SF}}\left(\frac{F_{{}_{\rm SF}}}{F_{\rm ch}}-1\right)
+ \frac{\lambda}{3}Q^{2}R_{\rm ch}^2\right],
\label{eq:delta_taylor2}
\end{align}
and the low-$Q^2$ expansion of the term in the square brackets
starts at the order $Q^4$ by construction. The nuclear models
\cite{Chen:2014sca,Chen:2014mza,RocaMaza:2012sj} are used
here---not to predict the distribution of weak charge in
${}^{12}$C, but rather---to determine the range of values
that need to be explored to quantify the uncertainty in $\Delta$.
These models, all informed by the charge radii and binding
energies of a variety of nuclei including ${}^{12}$C, predict
$|\lambda_{\rm SF}|\!\lesssim\!2\%$ with the central value of
$\lambda_{0}\!=\! - 0.90\%$.

\begin{figure}[t]
  \centering
  \resizebox{0.45\textwidth}{!}{\includegraphics{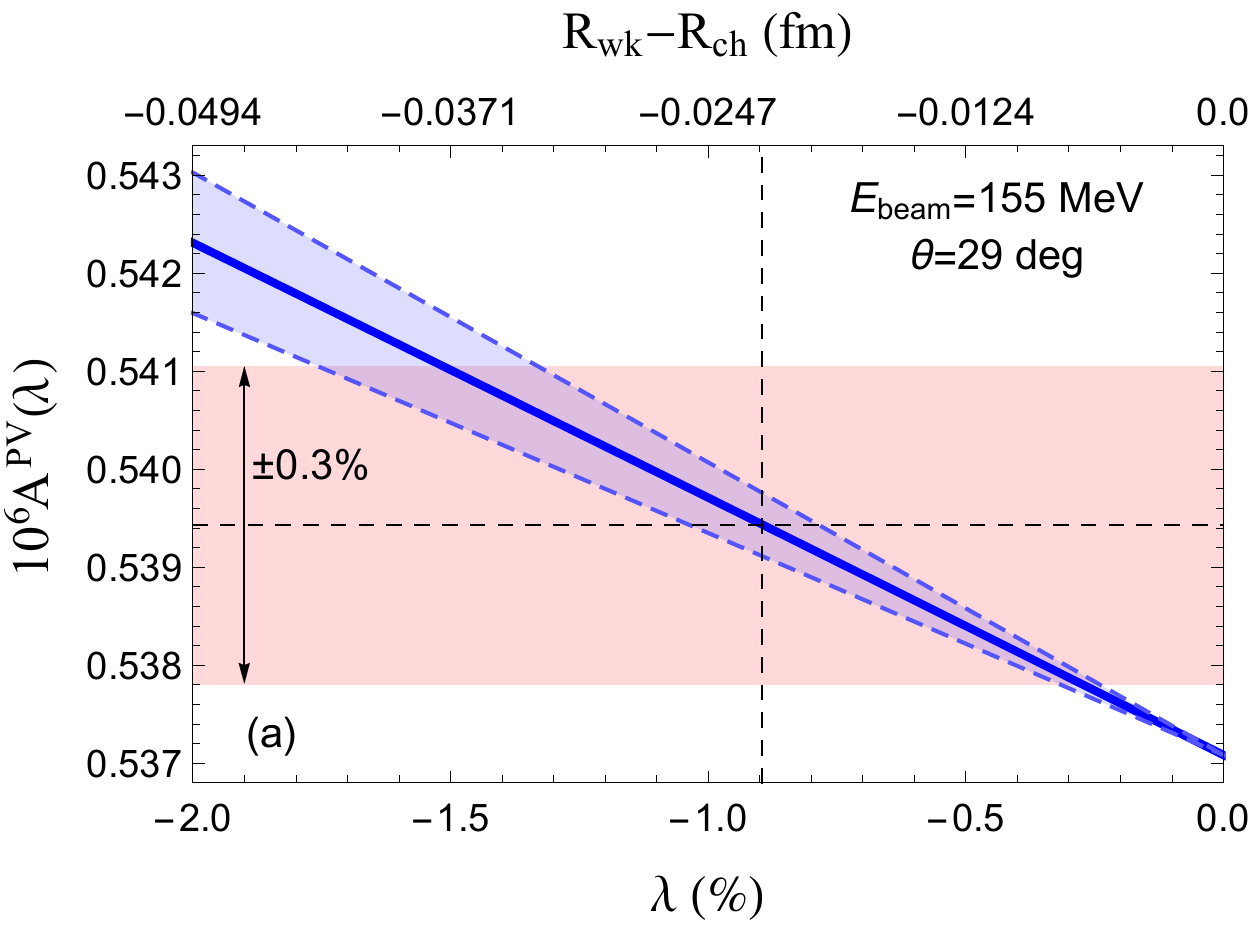}}
  \resizebox{0.45\textwidth}{!}{\includegraphics{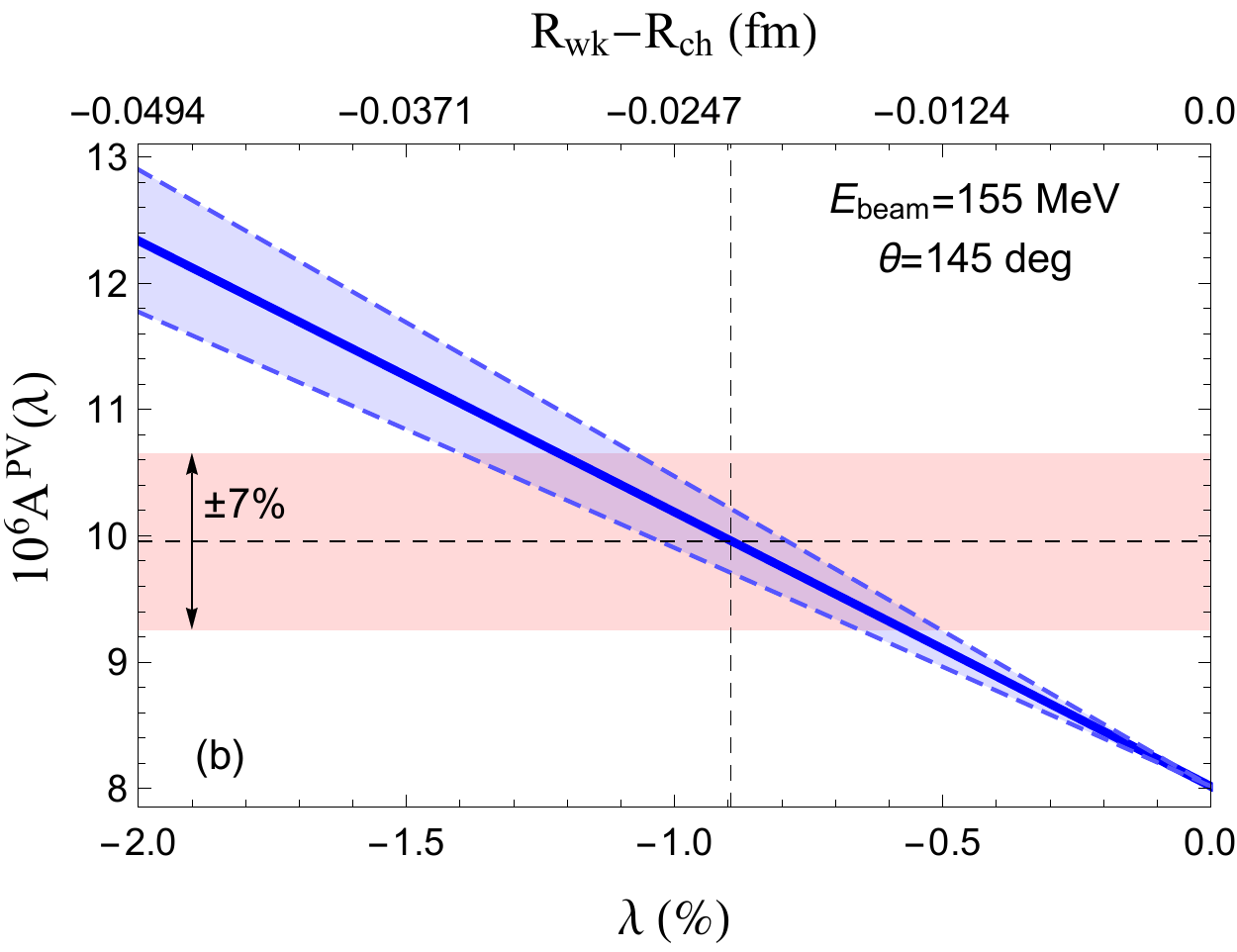}}
  \caption{
  The PV asymmetry at P2 with $E_{\rm beam}\!=\!155$ MeV for the
  (a) forward ($\theta\!=\!29^\circ$) and (b)
  backward measurement ($\theta\!=\!145^\circ$) as a
  function of the skin parameter $\lambda$. The pink-shaded band
  in the $\theta\!=\!29^\circ$ ($\theta\!=\!145^\circ$) plot
  indicates the anticipated (suggested) $0.3\%$ ($7\%$) precision
  in $A^{\rm PV}$. The blue band describes the residual model
  dependence as explained in the text. The vertical and horizontal dashed
  lines correspond to
  $\lambda\!=\!\lambda_0$ and $A^{\rm PV} (\lambda)\!=\!A^{\rm PV} (\lambda_0)$,
  respectively.
}
  \label{fig:Avslambda}
\end{figure}

To address the possibility to determine the weak charge of
${}^{12}$C with a precision of 0.3\% in the P2 experiment,
we study the sensitivity of the PV asymmetry to nuclear-structure
uncertainties. In Fig.~\ref{fig:Avslambda}, we display results
for $A^{\rm PV}$ as a function of $\lambda$ for an incident
electron energy of $E_{\rm beam}\!=\!155$\,MeV and two fixed
scattering angles: $\theta\!=\!29^\circ$ (upper panel) and
$\theta\!=\!145^\circ$ (lower panel). The blue band corresponds
to $\lambda$ varying in the range suggested by the models. Its
width indicates the spread of model predictions for $F_{\rm wk}$
and the central line is the bisector of the predictions for the
extreme choices of the models. The pink-shaded band in the
$\theta\!=\!29^\circ$ plot indicates
the anticipated $0.3\%$ precision in $A^{\rm PV}$.
From the sensitivity of the forward angle measurement to $\lambda$
shown in Fig.~\ref{fig:Avslambda}, one concludes that $\lambda$
should be known with a precision of 0.6\% or better to
constrain the weak charge of ${}^{12}$C to about 0.3\%.
Given that the nuclear models suggest a larger uncertainty in
$\lambda$, we conclude that with a single measurement and theory
input alone this task is not feasible.

Another option is to employ a second measurement of $A^{\rm PV}$
at $145^\circ$ (the lower panel of Fig.~\ref{fig:Avslambda})
to constrain the value of $\lambda$ to a narrower range. It is
seen that varying $\lambda$ in the adopted range translates into
a $\pm24\%$ variation in the asymmetry. Hence, a measurement
at this backward kinematical setting with a higher precision will
reduce the range of values of $\lambda$ and ultimately guarantee
a precise extraction of the weak mixing angle from a combination
of the two measurements. To a very good approximation the
$\lambda$-dependence of $A^{\rm PV}$ seen in Fig.\
\ref{fig:Avslambda} is linear, and we can write
\beqn
A^{\rm PV}
=
- \frac{G_FQ^2}{4\sqrt2\pi\alpha} \frac{Q_W}{Z} \Big(1+ p_0 + (p_1 + p_2 \zeta) \lambda \Big) ,
\label{eq:Deltapars}
\eeqn
so that the effect of varying $\zeta$ is depicted by the blue
bands in Fig.~\ref{fig:Avslambda}. The parameter $\zeta$ can be
chosen in such a way that $\zeta = \zeta_0 = 0$ corresponds to
the central prediction and $\zeta = \pm 1$ to the upper and
lower limits of the error band. Results of the distorted-wave
calculation of the coefficients $p_0$, $p_1$, and $p_2$
are provided in \hyperref[sec:supplemental]{Supplemental Material}.

\begin{figure}[t]
  \centering
  \resizebox{0.38\textwidth}{!}{\includegraphics{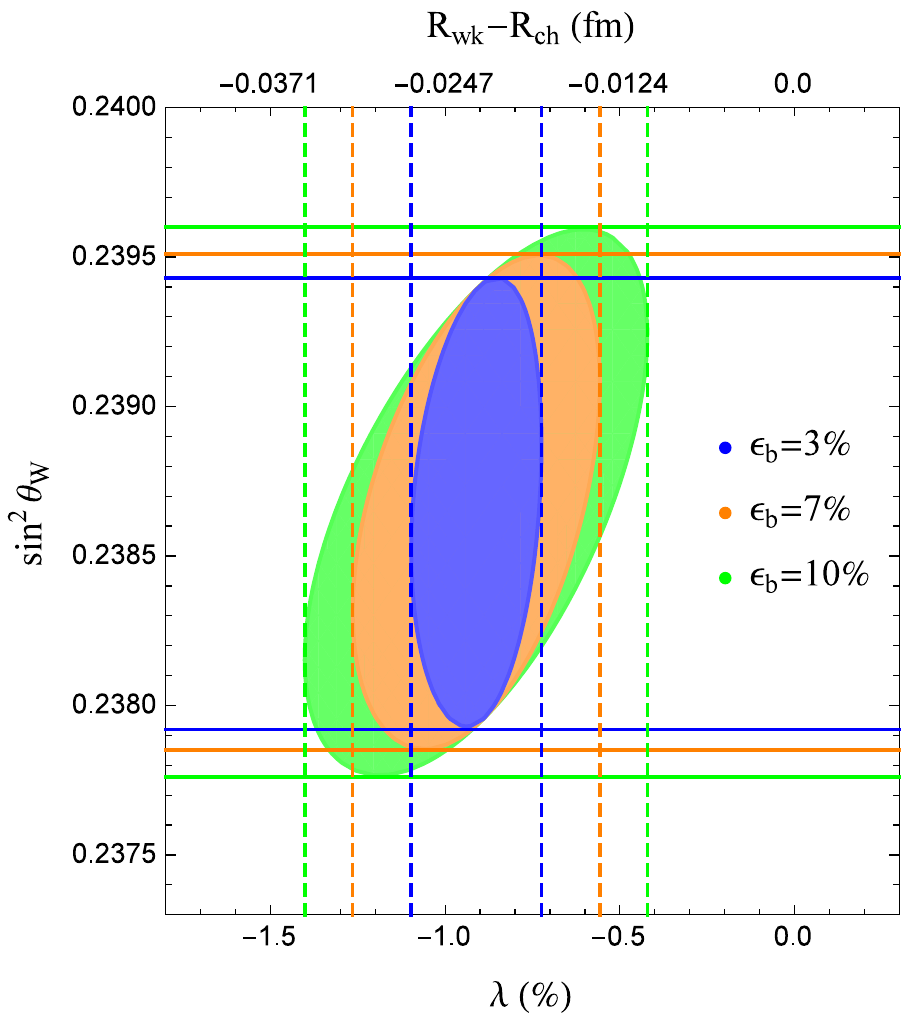}}
  \caption{
  The $1\sigma$-error regions from a fit of two expected
  measurements at forward and backward angles to the weak
  mixing angle, the skin parameter $\lambda$ and the model
  parameter $\zeta$. The plot shows the projection onto the
  $\sin^2\theta_W$-$\lambda$-plane. The green, orange and
  blue ellipses correspond to three assumptions for the
  precision $\epsilon_{\rm b}$ of the backward measurement
  (10\%, 7\%, and 3\%) whereas the precision of the forward
  measurement was always assumed to be 0.3\%.
  }
  \label{fig:chi2fit}
\end{figure}

We perform a $\chi^2$ fit of the combined forward ($A^{\rm PV}_{f}$)
and backward ($A^{\rm PV}_{b}$) measurements with respect to the
three free parameters $\sin^2\!\theta_W$, $\lambda$ and
$\zeta$. That is,
\begin{align}
\label{eq:chi2}
& \chi^2(\sin^2\!\theta_W, \lambda, \zeta) =
\\ &
\sum_{i=f,b}
\left(
\frac{A_i^{\rm exp} - A_i^{\rm PV}(\sin^2\!\theta_W, \lambda, \zeta)}
{\delta A_i} \right)^2
+ \left(\frac{\zeta - \zeta_0}{\delta \zeta}\right)^2
\, .
\nonumber
\end{align}
We assume that the experimental values of $A_i^{\rm exp}$ agree with
the SM prediction for which we choose the central value of
$\lambda\!=\!\lambda_0$. The experimental uncertainties
are given by $\delta A_i$. The last term in Eq.~(\ref{eq:chi2})
encodes our biases for the expected range of values of
$\zeta$, and we have chosen $\delta \zeta\!=\!1$. The $1\sigma$-allowed
range for $\sin^2\!\theta_W$ and $\lambda$ is obtained by solving
$\chi^2(\sin^2\theta_W, \lambda, \zeta)\!=\!1$. In
Fig.~\ref{fig:chi2fit} we show the projection of the
$\chi^2\!=\!1$ solution onto the $\sin^2\!\theta_W$-$\lambda$ plane
for three different choices of the precision of the backward-angle
measurement. The accuracy of the forward measurement remains fixed
at 0.3\%.
The covariance ellipses in Fig.~\ref{fig:chi2fit} suggest that
$\sin^2\!\theta_W$ and $\lambda$ are correlated and their
correlation decreases with increasing accuracy of the backward
measurement. Moreover, fractional uncertainties are given by:
$\delta\sin^2\theta_W/\sin^2\!\theta_W\!=\!\pm 0.39\%$ for
$\delta A_b/A^{\rm PV}_b\!=\!10\%$ and
$\delta\sin^2\theta_W/\sin^2\theta_W\!=\!\pm 0.35\%$ for
$\delta A_b/A^{\rm PV}_b\!=\!7\%$. An even higher precision
of the backward measurement, $3\%$, results in a reduction in
the uncertainty of the weak mixing angle to
$\delta\sin^2\!\theta_W/\sin^2\theta_W\!=\!\pm 0.32\%$.
At this point the uncertainty starts being dominated by the
forward measurement, so further improvement to the
backward measurement has no impact on the precision
of $\sin^2\!\theta_W$.

To summarize, we presented an ambitious proposal for a simultaneous
sub percent determination of the weak charge and weak radius of
${}^{12}$C using parity-violating electron scattering at the
upcoming MESA facility in Mainz. We demonstrated that to take
full advantage of an unprecedented 0.3\% precision aimed for in
the forward kinematical setting of P2 \cite{Becker:2018ggl}, an
additional $3\!-\!7\%$ measurement at a backward angle of
145$^\circ$ will ensure a largely model-independent extraction
of $\sin^2\!\theta_W$ with a relative precision of
$0.32\!-\!0.35\%$ and determination of $R_{\rm wk}$ within
$0.19\!-\!0.35\%$ of $R_{\rm ch}$.
Note that a similar combination of forward and
backward measurements on the proton is planned as part of the
P2 experiment \cite{Becker:2018ggl}, which makes the proposal
presented in this Rapid Communication a viable and attractive
possibility.

Whereas the weak skin of ${}^{12}$C and other symmetric nuclei
does not constrain the nuclear EOS, its exact value will
help quantifying generic isospin symmetry-breaking (ISB) effects.
Coulomb repulsion among the protons inside a nucleus and other
ISB mechanisms lead to a mismatch in the distribution of neutrons
and protons therein. Along with generating the proton (and weak)
skin of symmetric nuclei, ISB contributions play a major role in
the analysis of superallowed nuclear $\beta$ decays and the
extraction of $V_{ud}$ \cite{Hardy:2014qxa}. Importantly, in all
pairs of nuclei involved in the known superallowed $\beta$ transitions,
either the parent or the daughter nucleus is symmetric.
Therefore, precise information on weak skins of the (nearly)
symmetric parent and daughter nuclei will have an impact on the
tests of unitarity of the Cabibbo-Kobayashi-Maskawa matrix
and new physics searches with superallowed nuclear $\beta$ decays.

\begin{acknowledgments}
{\bf Acknowledgments} --
The authors acknowledge useful discussions with K.~Kumar and
F.~Maas. We thank T.~W.~Donnelly, R.~Ferro-Hern\'{a}ndez, and
O.~Moreno for helpful correspondence.
The work of J.~E., M.~G., O.~K., and H.~S.\ was supported by the
German-Mexican Research Collaboration Grants No.\ 278017
(CONACyT) and No.\ SP 778/4-1 (DFG).
M.~G.\ was supported by the EU Horizon 2020 Research and Innovation
Programme, Project No. STRONG-2020, Grant Agreement No.\ 824093.
X.~R-M.\ acknowledges funding from the EU Horizon 2020 research
and innovation programme, grant agreement No.\ 654002. J.~P.
acknowledges support from the U.S. Department of Energy
Office of Nuclear Physics under Award No. DE-FG02-92ER40750.
C.-Y.~S.\ was supported, in part, by the DFG (Grant No. TRR110)
and the NSFC (Grant No. 11621131001) through the funds provided
to the Sino-German CRC 110 ``Symmetries and the Emergence of
Structure in QCD,'' and by the Alexander von Humboldt
Foundation through the Humboldt Research Fellowship.
C.~J.~H. was supported by the U. S. Department of Energy Grants
No. DE-FG02-87ER40365 and No. DE-SC0018083.
\end{acknowledgments}

\begin{widetext}
\section{Supplemental Material}
\label{sec:supplemental}
1. In Table \ref{tab:2} we provide parameters of nuclear models
used in our calculation of the parity-violating asymmetry and
parameters $\zeta_f$ ($\zeta_b$) determined as a result of the
calculation which accounts for Coulomb distortion effects at
forward (backward) scattering angles.

\renewcommand{\arraystretch}{2}
\begin{table}[h]
\centering
\caption{\label{tab:2}}
\begin{tabular}{|l||*{7}{l|}}\hline
{Model} &
{$c, \mathrm{fm}$} &
{$a, \mathrm{fm}$} &
{$R_{\rm SF}, \mathrm{fm}$} &
{$\lambda_{\rm SF}, \%$} &
{$\zeta_f$} &
{$\zeta_b$} \\ \hline
RMF016 \ & $2.06065$ & $0.49389$ & $2.43274$ & $-1.52$ & $-1.00$ & $-1.00$ \\ \hline
RMF022 \ & $2.06849$ & $0.49445$ & $2.43830$ & $-1.29$ & $-0.80$ & $-0.30$ \\ \hline
RMF028 \ & $2.07585$ & $0.49544$ & $2.44482$ & $-1.03$ & $-0.44$ & $+1.00$ \\ \hline
RMF032 \ & $2.06421$ & $0.49433$ & $2.43578$ & $-1.39$ & $-0.89$ & $-0.62$ \\ \hline
SMC12 \    & $2.22693$ & $0.47318$ & $2.46358$ & $-0.27$ & $+1.00$ & $+0.59$ \\ \hline
\end{tabular}
\end{table}

The first four models listed in Table \ref{tab:2} fall under the general rubric of covariant (or
relativistic) energy density functionals. The models are based on an underlying Lagrangian
density that includes nucleons interacting via the exchange of various mesons and the
photon. In addition, nonlinear meson interactions are included to account for many-body
forces. The calibration of the handful of model parameters is informed by ground-state properties
of finite nuclei, their collective response, and constraints on the maximum neutron-star
mass\,\cite{Chen:2014sca}. Incorporated in the ground state properties are charge radii of
a variety of magic and semi-magic nuclei, including ${}^{12}$C. The outcome of the
calibration procedure is an optimal set of parameters together with a covariance matrix
that properly accounts for statistical uncertainties and correlations. The fitting protocol for
all the models is identical save one important distinction: the assumed value for the yet
to be determined neutron skin thickness of ${}^{208}$Pb ($R_{\rm skin}^{208}$).
Indeed, the neutron skin thickness of ${}^{208}$Pb is allowed to vary over the range of
$R_{\rm skin}^{208}\!=\!(0.16\!-\!0.32)\,{\rm fm}$\,\cite{Chen:2014mza}.

The model named SMC12 is a non-relativistic energy density functional of the Skyrme type. SMC12 has been devised to reproduce the binding energy ($B$) and charge radii ($R_{\rm ch}$) of ${}^{12}$C without compromising the accuracy in the description of other observables along the nuclear chart. Specifically, the fitting protocol has been based on that of the SAMi interaction \cite{RocaMaza:2012sj} with the following modifications: i) inclusion of ${}^{12}$C data ($B$ and $R_{\rm ch}$); and ii) relaxation of the weight on the pure neutron matter equation of state. This allowed us to accommodate the new data within the presented model. As an example, the experimental charge radii and nuclear masses of ${}^{12}$C, ${}^{16}$O, ${}^{40}$Ca, ${}^{48}$Ca, ${}^{90}$Zr, and ${}^{208}$Pb are reproduced to better than 1\% accuracy, except for the binding energy of the two Calcium isotopes which are accurate at the percent level. This example justifies the reliability of the model for the present study.

To sum up, the five models that we selected for our analysis cover a fairly wide
landscape in the parameter space. For the energy density functionals that we used,
the predicted neutron skin thickness of $^{208}$Pb ranges from about $0.12\,{\rm fm}$
all the way to $0.32\,{\rm fm}$; going beyond this range is hard without compromising
the agreement with experiment.

2. In Fig. \ref{fig:1} we present an example of the calculation
which accounts for Coulomb distortion effects at forward ($\theta\!=\!29^\circ$)
and backward ($\theta\!=\!145^\circ$) scattering angles.
\begin{figure}[h]
    \centering
    \resizebox{0.38\textwidth}{!}{\includegraphics{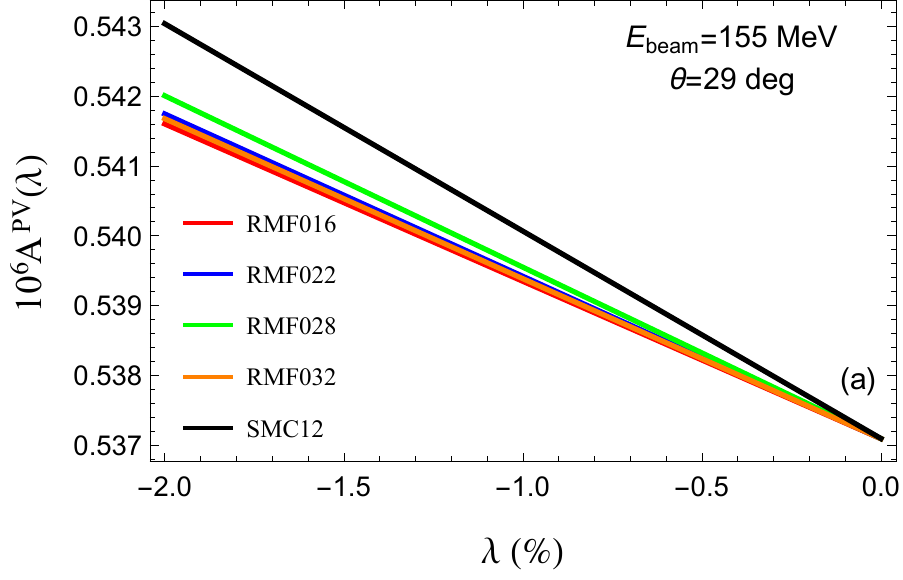}}
    \resizebox{0.362\textwidth}{!}{\includegraphics{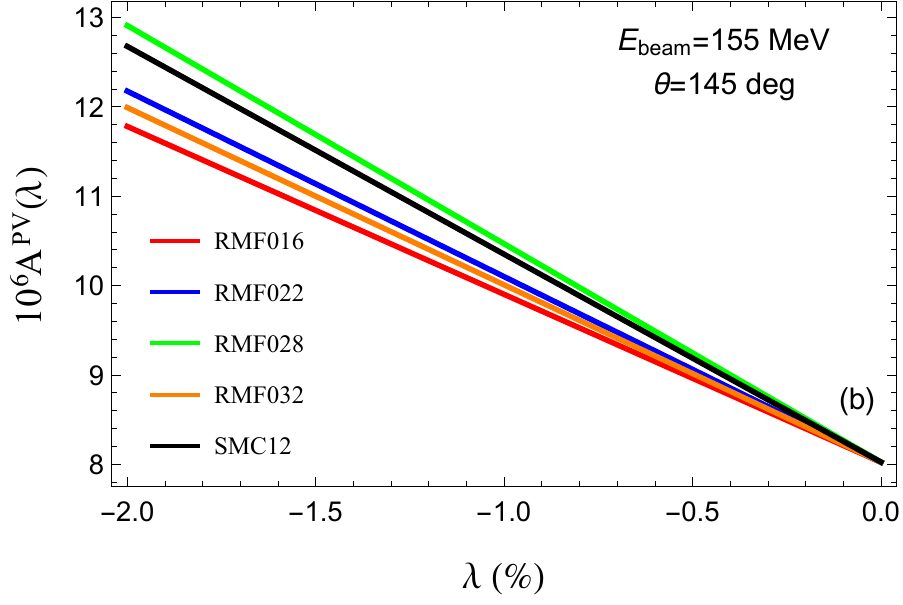}}
    \caption{\label{fig:1} Predictions for $A^{\rm PV}$
    at (a) $\theta\!=\!29^\circ$ and (b) $\theta\!=\!145^\circ$.}
\end{figure}

3. In Table \ref{tab:1} we provide the coefficients $p_0$, $p_1$, and $p_2$
obtained as a result of the calculation which accounts for Coulomb
distortion effects at forward and backward scattering angles.
These coefficients are defined by Eq.~(\ref{eq:Deltapars}).
\renewcommand{\arraystretch}{2}
\begin{table}[h]
\caption{\label{tab:1}}
\begin{center}
\begin{tabular}{|l||*{3}{l|}}\hline
{Coefficient} &
{$\theta\!=\!29^\circ$} &
{$\theta\!=\!145^\circ$} \\ \hline
$p_0$ &$+0.04005$ &$+0.09586$ \\ \hline
$p_1$ &$-0.50612$ &$-29.5132$ \\ \hline
$p_2$ &$-0.06969$ &$-3.86420$ \\ \hline
\end{tabular}
\end{center}
\end{table}
\end{widetext}


\begin{thebibliography}{9}
\bibitem{ftnt1}
The discussion of electroweak radiative corrections is deferred
to a future work.
\bibitem{ftnt2}
Other choices of the reference frame would introduce corrections
depending on the nuclear recoil. Up to now, no fully consistent
way of accounting for recoil corrections in the Dirac formalism
is known.
\end{thebibliography}
\end{document}